\documentclass[showpacs,showkeys,preprintnumbers,aps,amsmath,amssymb]{revtex4}
\usepackage{graphicx}
\usepackage{dcolumn}
\usepackage{bm}

\def\bfmu{{\bf \mu}}

\begin{document}
\title{A physical application of $g$-function}
\author{Sang-Hoon Kim}
\affiliation{Division of Liberal Arts, Mokpo National Maritime
University, Mokpo 530-729, R. O. Korea}
\email{shkim@mmu.ac.kr}
\date{\today}
\begin{abstract}
We investigated an integral representation of a complex function
which was obtained from the molecular beam magnetic resonance, and
named it as `$g$-function' because it was connected with Gamma
function as a special case. We  introduced a physical example of the
$g$-function from a dipole-dipole interaction of rigid polar
molecules.
\end{abstract}
\pacs{02.30.Sa, 02.30.Gp, 03.50.De}
\keywords{Gamma function,
complex function, dipole-dipole interaction }
\maketitle


Quantum mechanics cannot stand without special functions such as
Bessel function or Legendre function. Besides these functions, there
are so many functions that was originated from mathematics but has
been used in physical sciences more than mathematics. On the other
hand there are still many functions that has been studied by
mathematician but has been rarely used in other areas. Among them
some functions do not have even names.

In this paper,
 we will introduce a mathematical function which has not name yet
and has been considered as no physical application. We call it as
$g$-function because the gamma function is a very special case of
the function. Of course it is totally different from the generalized
or modified Gamma functions. We'll use small `$g$' to avoid the
confusion from the Meijer's $G$-function \cite{gra}. Then, we will
introduce a physical example of the typical case of the
$g$-function.
 from a familiar classical model of an electromagnetic theory.

In various theoretical physics such as a neutron beam velocity
distribution function \cite{zah,lap}  or a theory of  the molecular
beam magnetic
 resonance method \cite{rab,tor,ram},
 there arise a definite integral function
 $\int_0^\infty  y e^{-y^2 - z/y} dy$,
where $z$ is a complex number. It was the birth of a new complex
function
 containing the Gamma function
as a special case.

In 1937 the $g$-function was introduced by Zahn \cite{zah} from an
absorption coefficients for slow thermal neutrons. The fraction of
neutrons transmitted is given by
\begin{equation}
\phi_{1}(x) = \int_0^\infty y e^{-y-x/y^{1/2}} dy. \label{10}
\end{equation}

Then, the function was extended to $m$-th order
 by Laporte \cite{lap}. The extended form was given by
\begin{equation}
\phi_{m}(x) = \int_0^\infty y^{m} e^{-y-x/y^{1/2}} dy. \label{12}
\end{equation}
The order $m$ is restricted to positive integer here. He showed that
the integral function is a solution of a third-order linear
differential equations. Also the convergent expansions were obtained
for small values of $x$.

Later in 1941 the modern form of the $m$-th order $g$-function was
introduced by Torrey \cite{tor}
 in a neutron beam velocity distribution.
The transition probability in radio-frequency spectra contains the
function. At result in 1951 the 3rd order $g$-function was  studied
mathematically by Kruse and Ramsey \cite{kru}. At this time the
$g$-function was extended into complex plane.

The complete form of the $m$-th order $g$-function is defined by
\begin{equation}
g_m(z) = \int_0^\infty  y^m e^{-y^2 - z/y} dy, \label{13}
\end{equation}
where $m=0, 1, 2, ...$, and $z$ is a complex number. Interchanging
$y$ into $1/x$, we obtain another expression of $g_m(z)$ as
\begin{equation}
g_m(z) = \int_0^\infty  \frac{e^{-1/x^2 - zx}}{x^{m+2}} dx.
\label{15}
\end{equation}
Therefore, the Gamma function is a special case of the $g$-function
as
\begin{equation}
g_m(0) = \frac{1}{2}\Gamma\left(\frac{m+1}{2}\right). \label{17}
\end{equation}
Alternatively,
\begin{equation}
g_{2m+1}(0) = \frac{n!}{2}. \label{18}
\end{equation}


The  $g_m$ has the following three fundamental properties
\cite{abr}. First, it is a solution of the 3rd order differential
equation
\begin{equation}
x \frac{d^3 g_m}{dx^3} - (m-1)\frac{d^2 g_m}{dx^2} + 2 g_m (x) = 0,
\label{20}
\end{equation}
where $m=0, 1, 2, ...$ Second, the derivative reduces single order
and change its sign
\begin{equation}
 \frac{d g_m}{dx} =  - g_{m-1}(x),
\label{25}
\end{equation}
where $m=1, 2, 3,...$. Third, it satisfies the recurrence relation
\begin{equation}
2 g_m(x) = (m-1)g_{m-2}(x) + x g_{m-3}(x), \label{30}
\end{equation}
where $m=3, 4, 5,...$. More basic properties of the $g_m$ in pure
mathematical aspects are summarized in the handbook by Abramowitz
and Stegun \cite{abr}.

The $g_m(z)$ can be also written as $g_m(r,\theta)$ in the polar
coordinate. The complex $z$ is written as
 $z=r e^{-i\theta}$,
where $r=|z|$ and $0 \le \theta \le \pi/2$. Then, the real and
imaginary parts are
\begin{equation}
Re g_m(r,\theta) =\int_0^\infty y^m e^{-y^2-\frac{r\cos\theta}{y}}
\cos\left(\frac{r\sin\theta}{y}\right) dy, \label{125}
\end{equation}
and
\begin{equation}
Im g_m(r,\theta) =\int_0^\infty y^m e^{-y^2-\frac{r\cos\theta}{y}}
\sin\left(\frac{r\sin\theta}{y}\right) dy. \label{130}
\end{equation}
It oscillates very slowly around $r$ axis but converges to zero
 quickly.

Furthermore, $g_m(z)$, $m=0, 1, 2, ...$,
 does not have infinite Taylor series expansion because
it is expanded up to the $m$-th order of $z$ at $z=0$. From Eqs.
(\ref{15}) and (\ref{25}) the $n$-th derivatives of $g_m(z)$ by $z$
is
\begin{equation}
\frac{d^n g_m(z)}{dz^n} = (-1)^n \int_0^\infty
 \frac{e^{-1/x^2 - zx}}{x^{m-n+2}} dx.
\label{53}
\end{equation}
where $n \ge 1$. If we check the value at the origin,
\begin{eqnarray}
\lim_{|z|\rightarrow 0} \frac{d^n g_m(z)}{dz^n} &=& {\rm finite \,
constant :}
 \hspace{.1in} n \le m.
\nonumber \\
&=& {\rm infinite :} \hspace{.1in} n > m. \label{120}
\end{eqnarray}

Although it has some possibilities of exciting features,
 there has been no more serious studies in the $g_m$ function
  since 1950s.
The main reason is that the application area of the function
 has never been found besides the previous known fields.
Furthermore, the function was not considered seriously as an
extended form of the Gamma function. That is, the further studies of
the $g_m$ was retarded
 by the restriction of the application area
even though it has many interesting properties.


We introduce a classical example of the $g_1(z)$ in a complex plane.
It comes from the Fourier transformation of the time correlation
function of dipoles in linear response theory. We begin our story of
the function from an example: the frequency dependent electric
susceptibility of polar molecules.

We assume a homogeneous isotropic fluid of rigid  particles. In
general the shape of a rigid molecule is arbitrary, and has three
different moment of inertia. Also we assume that it has a dipole
moment $\bfmu(t)$ in the direction of rotating axis. Let $\omega$ be
the applied frequency and $\tau$ be a relaxation
 time by collision of the rotating particles.
Then, the self part of the electric susceptibility has the following
form \cite{kub,han,wei,kim}.
\begin{equation}
\frac{\chi_s (\omega+i/\tau )}{\chi_s(0)}
 = 1 + i (\omega+i/\tau)
\int_0^\infty e^{i(\omega+i/\tau)t} \frac{\langle {\bf \mu}(0) \cdot
{\bf \mu}(t) \rangle }{\mu^2} dt. \label{35}
\end{equation}
$\langle...\rangle$ denotes an ensemble average over initial
conditions or the statistical average in the absence of applied
fields.

 Let $ (\theta_i(t),\phi_i(t),\psi_i(t))$ be
the Euler angles that describe the orientation of particle $i$ at
time $t$ with respect to a laboratory fixed frame of reference.
Also, let  $I_i$ and $ \Omega_i (i=1,2,3)$ be the moment of inertia
and angular velocity of the three dimensional rigid body to its
three principal axes. Then, the angular momentum $L_i$ satisfies $
L_i = I_i  \Omega_i$. The kinetic energy of an arbitrary shape of
rotating three dimensional rigid body is composed of translational
and rotational motion. That is,
\begin{eqnarray}
E&=& \frac{ P^2}{2M}+ \frac{1}{2}\sum_{i=1}^3 I_i \Omega_i^2
\nonumber \\
&=& \frac{ P^2}{2M}+ \frac{f(\theta,\psi)  L^2}{2 I_3}, \label{40}
\end{eqnarray}
where $M$ is the mass and $ P$ is the linear momentum. The
$f(\theta,\psi)$ is the angular momentum relation function that is
defined by
\begin{equation}
f(\theta,\psi) = \frac{I_3}{I_1}\sin^2\theta \sin^2\psi +
\frac{I_3}{I_2}\sin^2\theta \cos^2\psi + \cos^2\theta. \label{45}
\end{equation}
 $f(\theta,\psi)$ varies {\it smoothly} in $\theta$ and $\psi$,
and it gives us the relation between the rotational kinetic energy
 and  the angular momentum.
The translational motion of the particle gives just a constant in
calculation of the statistical average. Therefore, we can neglect $
P^2/2M$ term in further steps.

For convenience we define the angle difference of the dipole ${\bf
\mu}$ at time $0$ and $t$ as $\gamma$.
 Then, we can write the dipole-dipole correlation function
in Eq. (\ref{35}) as
\begin{equation}
\frac{ {\bf \mu}(0) \cdot
 {\bf \mu}(t)} {\mu^2}=\cos\gamma(\theta,\psi,u),
\label{50}
\end{equation}
where $u$ is a dimensionless time defined by $u = Lt/I_3$. This
conversion makes the $\cos\gamma$ not include $L$ explicitly.

\begin{figure}
 \includegraphics{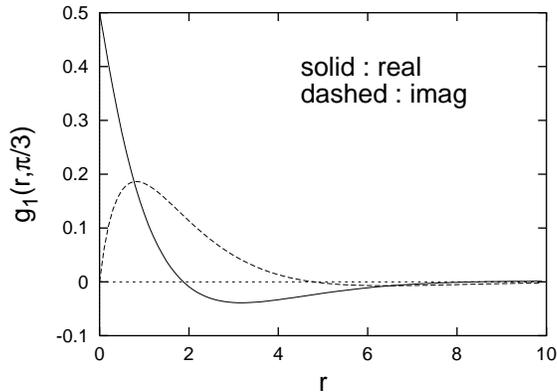}\\
\caption{The radial dependence of $g_1(r,\pi/3)$.
}
\end{figure}

\begin{figure}
 \includegraphics{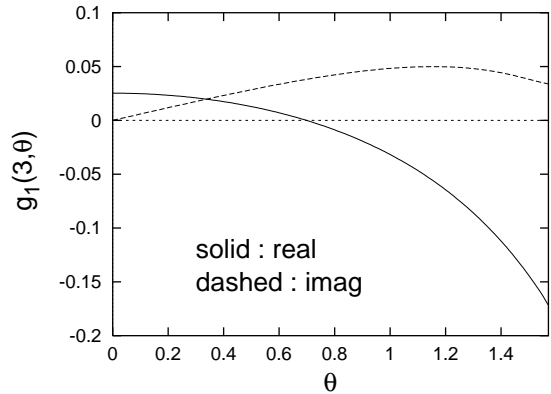}\\
\caption{The phase dependence of  $g_1(3,\theta)$. }
\end{figure}

Applying  Eqs. (\ref{40}) and (\ref{50}) to  Eq. (\ref{35}) produces
$g(\theta,\psi,u)$ as an
 essential part of the Fourier transformation.
\begin{equation}
\frac{\chi_s (\omega+i/\tau )}{\chi_s(0)} =1 + \frac{i
(\omega+i/\tau)}{Z} \int_0^\infty du \int_0^\pi  \sin\theta \,
d\theta \int_0^{2\pi} \cos\gamma(\theta,\psi,u) g(\theta,\psi,u)
d\psi, \label{60}
\end{equation}
where
\begin{equation}
g(\theta,\psi,u) = \int_0^\infty  L e^{-(\beta f /2I_3) L^2 -
(1/\tau-i\omega)I_3 u/L} dL. \label{65}
\end{equation}
The  $Z$ is the partition function and independent of the initial
position and time.
 $\beta=1/k_B T$,
where  $k_B$ is the Boltzmann constant
 and $T$ is the absolute temperature.

Eventually, the integral expression of $g(\theta,\psi,u)$
 is directly related with the $g_1(z)$ in complex plane as
\begin{equation}
g_1(z) = \int_0^\infty  y e^{-y^2 - z/y} dy. \label{70}
\end{equation}
The $z$ is a complex number given by
 $z=z_1 - i z_2$.
Comparing Eq. (\ref{60}) with  Eq. (\ref{65}), the regions of the
real number $z_1$ and $z_2$ are decided as
 $z_1>0,$ and $ z_2 >0$.
We plotted the real and imaginary parts  of the $g_1(r,\theta)$. The
radial dependence at $\theta=\pi/3$ was plotted in FIG. 1, and we
see that it oscillates and quickly drops to zero. The phase
dependence at $r=3$ is plotted in FIG. 2, too.


We named a special function as $g_m$ function with order $m$, and
studied some fundamental properties of it. A physical example of the
$g$-function in a complex plane was introduced from classical
mechanics and electromagnetic theory.

Although, some functions has been regarded as not useful application
areas, it is not actually true. We have not found the application
area yet. It is interesting enough to study connecting with the
Gamma function. Merging to 0 of the radial part is
 quicker than Bessel functions and Neumann functions,
and that may applicable to a new integral transformation. We hope
that it  covers various range of physics that previous ordinary
Gamma function could not.



\begin{thebibliography}{}
\bibitem{gra} I.S. Gradshteyn and I.M. Ryzhik,
    {\it Tables of Integrals, Series, and Products}, 4th,
    (Academic, New York, 1965).
\bibitem{zah} C.T. Zahn, Phys. Rev. {\bf 52}, 67 (1937).
\bibitem{lap} O. Laporte, Phys. Rev. {\bf 52}, 72 (1937).
\bibitem{rab} I.I. Rabi, Phys. Rev. {\bf 51}, 652 (1937).
\bibitem{tor} H.C. Torrey, Phys. Rev. {\bf 59}, 293 (1941).
\bibitem{ram} N.F. Ramsey, Phys. Rev. {\bf 78}, 695 (1950).
\bibitem{kru} U.E. Kruse and N.F. Ramsey,
    J. Mathematics and Physics, {\bf 30}, 40 (1951).
\bibitem{abr} M. Abramowitz and I.A. Stegun, edited,
    {\it Handbook of Mathematical Functions with Formulas, Graphs,
    and Mathematical Tables}, (Dover, New York, 1970).
\bibitem{kub} R. Kubo, {\it Statistical Mechanics}
    (North Holland, Amsterdam, 1964)
\bibitem{han} J.-P. Hansen and I.R. McDonald,
    {\it Theory of Simple Liquids}, 2nd, (Academic, London, 1990).
\bibitem{wei} D. Wei and G.N. Patey, J. Chem. Phys. {\bf 91},
    7113 (1989); {\it ibid.} {\bf 93}, 1339 (1990).
\bibitem{kim} S.-H. Kim, G. Vignale, and B. DeFacio, \pra
    {\bf 46}, 7548 (1992); S.-H. Kim, G. Vignale, and B. DeFacio,
    \pre {\bf 50} 4618 (1994).
\end{thebibliography}
\end{document}